\documentclass[a4paper,12pt]{article}
\usepackage{latexsym}
\usepackage{amssymb}
\usepackage{amsmath}
\usepackage[dvips]{graphicx}
\usepackage{cite}
\def\dst{\displaystyle}
\begin{document}
\title{\bf Inclusive distributions in the unitarized pomeron models}
\author{A. Alkin$^{(1)}$, E. Martynov$^{(1)}$,  V. Pauk$^{(2)}$, E. Romanets$^{(2)}$\\
~\\
\leftline{$^1$ \small Bogolyubov Institute for Theoretical Physics,
{\small \it~~Metrologichna 14b, Kiev, UA-03680, Ukraine}}\\
\leftline{$^2$ \small Taras Shevchenko Kiev National University, 
\leftline{\small \it~~Volodimirska 60, Kiev, UA-03101, Ukraine.}} \\
\\}
\maketitle

\begin{abstract}
High energy inclusive hadron production in the central kinematical region is analyzed within the models of unitarized pomeron. It is shown that the sum of multipomeron exchanges with intercept $\alpha_P(0)>1$ reproduce qualitatively  contribution of the triple pole (at $t=0$) pomeron to inclusive cross section. Basing on this analogy we then suggest a general form of unitarized pomeron contributions (in particular the dipole or tripole pomeron) to inclusive cross section. They lead to a parabolic form of the rapidity distribution giving $<n>\propto \ln^3s$ (tripole) or $<n>\propto \ln^2s$ (dipole). The models considered with suggested parametrization of $p_t$-dependence for cross sections well describe the rapidity distributions data in $pp$ and $\bar pp$ interactions at energy $\sqrt{s}\geq 200$ GeV. The predictions for one particle inclusive production at LHC energies are given.
\end{abstract}

Pomeron with intercept $\alpha_{P}(0)=1+\varepsilon,\quad \varepsilon >0$ is a very attractive model from the phenomenological point of view \cite{DonDoshLanNach}. It gives a simple and compact  parametrization for various  high-energy soft processes (elastic and deep inelastic scattering, diffraction and others) and describes quite well a lot of data for high enough energy (for example, total cross sections and small-$t$ ($\lesssim 1$ Gev$^{2}$) elastic scattering at $\sqrt{s}\geq 5$ GeV).

On the other hand at $s\to \infty$ the contribution of such a pomeron violates unitarity explicitly. The model leads  to total cross section of hadron interaction behaving as $\sigma_{t}(s)\propto (s/s_{0})^{\varepsilon}$ ($s_{0}=1$ GeV$^{2}$) in contradiction to the Heisenberg-Froissart-Martin-Lukaszuk theorem
\begin{equation}
\sigma_{t}\leq \frac{\pi}{m_{\pi}^{2}}\ln^{2}(s/s_{0}).
\end{equation}
Thus the model can be considered only as phenomenological tool and must be improved in a some manner in order to restore unitarity, at least, to avoid a rough violation of unitarity bound on the total cross section. Let us remind that originally the contribution of of such a pomeron corresponds to a simple pole of partial amplitude in the plane of complex angular momentum.

\begin{figure}
\begin{center}
\includegraphics[scale=0.7]{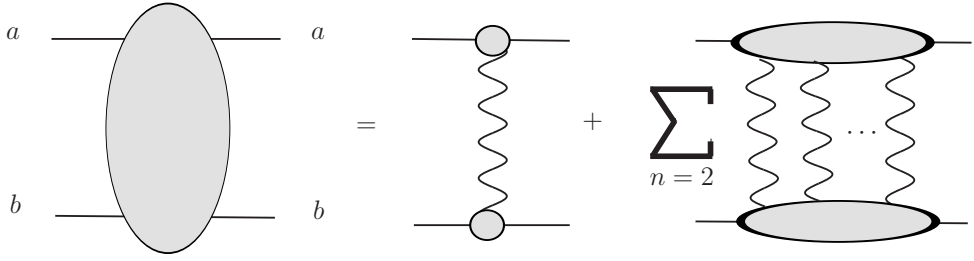}
\caption{Multipomeron contributions to elastic scattering amplitude}
\label{fig:rescatter}
\end{center}
\end{figure}

There are the several ways to restore unitarity. The most simple method to do that is to sum multipomeron diagrams up (Fig. \ref{fig:rescatter})

\begin{figure}[h]
\begin{center}
\includegraphics[scale=0.8]{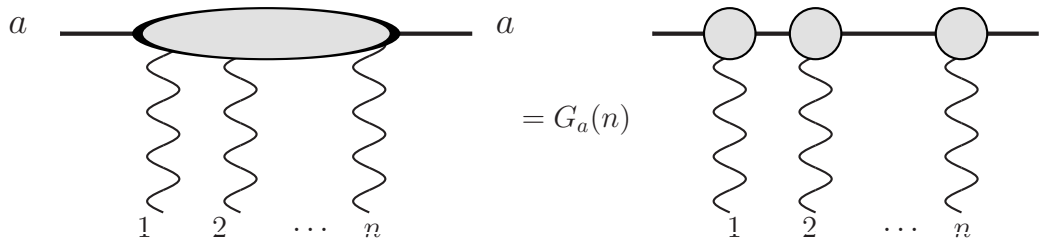}
\caption{Amplitude of interaction of two hadrons with $n$ pomerons in a pole approximation but with phenomenological factor $G_{a}(n)$.}\label{fig:2hadrnpom}
\end{center}
\end{figure}

Starting on one pomeron exchange which can be written in the form

\begin{equation}\label{eq:oneP}
a(s,t)=\eta_{P}(t)\tilde g_{ab}(t)\left(\frac{s}{s_{0}}\right)^{\alpha_{P}(t)}=
-g_{a}(t)g_{b}(t)\left(-i\frac{s}{s_{0}}\right)^{\alpha_{P}(t)}
\end{equation}
where $s_{0}=1 {\rm GeV^{2}}$ and
$$\eta_{P}(t)=\frac{1+\exp(-i\pi\alpha_{P}(t))}{-sin{\pi\alpha_{P}(t)}},$$
then going to the impact parameter representation
\begin{equation}\label{eq:impact}
h(s,b)=\frac{1}{8\pi s}\int\limits_{0}^{\infty}dbbJ_{0}(gb)a(s,-q^{2}),
\end{equation}
one can obtain under certing simplifying assumption (see below) the amplitudes
\begin{equation}\label{eq:Hsb}
H(s,b)=-\frac{1}{2i}\sum\limits_{n=1}^{\infty}\frac{G_{a}(n)G_{b}(n)}{n!}[-2ih(s,b)]^{n}.
\end{equation}
It appears in the form Eq.(\ref{eq:Hsb}) if we assume that two-hadrons-{\it n}-pomeron amplitude is proportional to the product of two-hadron-pomeron vertices (it is a pole approximation for intermediate states) as shown in Fig. \ref{fig:2hadrnpom}.
Moreover, assuming either $G(n)=C^{n}$ or $G(n)=C^{n}\sqrt{n!}$ we obtain from Eq.(\ref{eq:Hsb}) two well known schemes of pomeron unitarization: eikonal \cite{eikonal} or quasi-eikonal \cite{quasieik}  and quasi-$U$-matrix models \cite{U-matrix,quasiU}.
\begin{equation}\label{eq:QE&U}
H(s,b)=\left \{
\begin{array}{lll}
&\dst \frac{1}{2iC_{a}C_{b}}\left( 1-e^{-2iC_{a}C_{b}h(s,b)}\right), &\quad {\rm if} \quad G_{a,b}(n)=C_{a,b}^{n}\\
&&\\
&\dst \frac{h(s,b)}{1+2iC_{a}C_{b}h(s,b)}, &\quad {\rm if} \quad G_{a,b}(n)=C_{a,b}^{n}\sqrt{n!}
\end{array}
\right .
\end{equation}
If $\alpha_{P}(t)=1+\varepsilon +\alpha_{P}^{'}t$ and $g_{a,b}(t)=\exp(B_{a,b}t)$ one can find that at $s\to \infty$ in the both models
\begin{equation}\label{eq:sig-t}
\sigma_{t}^{ab}(s)\approx8\pi\varepsilon R^{2}(s)\ln(s/s_{0})\approx 8\pi\varepsilon\alpha_{P}^{'}\ln^{2}(s/s_{0}) ,
\end{equation}
where $R^{2}(s)=B_{a}+B_{b}+\alpha_{P}^{'}\ln(s/s_{0})$.

This result provides a ground for another method for constructing the amplitude. The possible way is to consider partial amplitudes with more complicated singularities than usual simple angular momentum poles just from the beginning. It is worth to emphasize that the factorization of residues is valid not only for the simple $j$-poles but also for any isolated $j$-singularity \cite{factorization} of amplitude. Thus one can consider, for instance, double pole (dipole pomeron) \cite{dipole} instead of simple pole. In this model $\sigma_{t}(s)\propto\ln(s/s_{0})$ at $s\to \infty$. Another interesting possibility comes from triple pole at $t=0$ (at $t\neq 0$ because of unitarity it must be a pair of colliding at $t=0$ hard branch points producing a triple pole) \cite{tripole}. Triple pomeron gives $\sigma_{t}(s)\propto\ln^{2}(s/s_{0})$. Both models lead not only to a very good description  of the hadron total cross sections but the differential elastic cross sections, deep inelastic scattering and vector meson photoproduction as well.

In the light of such a quite successful applications of unitarized pomeron models it is interesting to see how manyparticle processes can be described in this framework. We consider here one particle distribution in rapidity and pseudorapidity. In the next section we will remind some of the results obtained  earlier for  multipomeron exchanges. Then we suggest the specific parametrizations of $p_{t}$ and $y$ dependence in the dipole and tripole pomeron models, compare them with the availble data for $200 \leq \sqrt{s}\leq $ 1800 GeV and make predictions for future experiments at higher energy.

\section{Multipomeron exchanges in the model with $\alpha_{P}(0)>1$}
\begin{figure}[h]
\begin{center}
\includegraphics[scale=0.4]{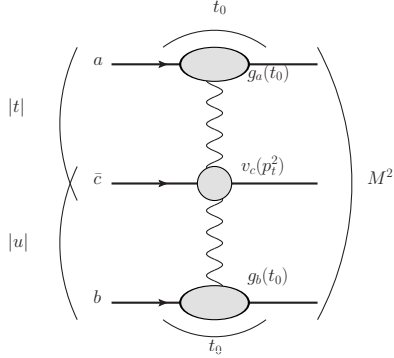}
\caption{Pomeron contribution to inclusive production in the central region}
\label{fig:one-Pom}
\end{center}
\end{figure}

Due to generalized optic theorem differential cross section of one particle inclusive production ($a+b\to c+X$) in the central kinematical region where
\begin{equation}\label{eq:kinematics1}
\begin{array}{llll}
s&=&(p_{a}+p_{b})^{2}\to \infty, &\\
t&=&(p_{a}-p_{c})^{2}, & |t|\to \infty,\\
u&=&(p_{b}-p_{c})^{2}, & |u|\to \infty,\\
M^{2}&=&(p_{a}+p_{b}-p_{c})^{2}, & s/M^{2}\to 1
\end{array}
\quad \frac{tu}{s}=m_{t}^{2}=m_{c}^{2}+p_{l}^{2}
\end{equation}
is related with the diagram of Fig. \ref{fig:one-Pom}.

More exactly, at large energy and for the simple pomeron pole with $\alpha_{P}(0)-1=\varepsilon$
\begin{equation}\label{eq:incl-c-s}
\begin{array}{ll}
\dst E\frac{d^{3}\sigma}{d^{3}p}=E\frac{d^{3}\sigma}{dp_{l}d^{2}p_{t}}=&8\pi Disc_{M^{2}}{\mathcal M}(a+b+\bar c\to a+b+\bar c)=\\&\dst g_{a}(0)\left (\frac{|t|}{s_{0}}\right)^{\varepsilon}v_{c}(p_{t}^{2})\left (\frac{|u|}{s_{0}}\right)^{\varepsilon}g_{b}(0).
\end{array}
\end{equation}
where $E,p_{l},\vec p_{t}$ are energy and momenta of the inclusive hadron $c$, $g_{a,b}(t_{0}=0)$ are the coupling vertices $aPa, bPb$, $s_{0}=1{\rm GeV}^{2}$.

It is more convenient for what follows to use another set of variables, $(p_{t},y)$ or $(p_{t},\eta)$,
\begin{equation}\label{eq:rapidities}
y=\frac{1}{2}\ln\frac{E+p_{l}}{E-p_{l}}, \quad \eta=-\ln(\tan\vartheta/2),
\end{equation}
where $\vartheta$ is the scattering angle of hadron $c$ in the center mass system. With a rapidity variable the cross section Eq.(\ref{eq:incl-c-s}) is read as
\begin{equation}\label{eq:incl-c-s-y}
E\frac{d^{3}\sigma}{d^{3}p}=g_{a}(0)e^{\varepsilon(y_{0}-y)}v_{c}(p_{t}^{2})e^{\varepsilon(y_{0}+y)}g_{b}(0)=
g_{a}(0)v_{c}(p_{t}^{2})g_{b}(0)e^{2\varepsilon y_{0}}.
\end{equation}
Here $y_{0}$ is the maximal value of rapidity $y$. In the c.m.s.
\begin{equation}\label{eq:lims-y}
-1/2\ln(s/m_{t}^{2})\leq y\leq 1/2\ln(s/m_{t}^{2}), \quad {\rm i.e.}\quad y_{0}=1/2\ln(s/m_{t}^{2}).
\end{equation}

\begin{figure}[h]
\begin{minipage}{7.cm}
\begin{center}
\includegraphics[scale=0.7]{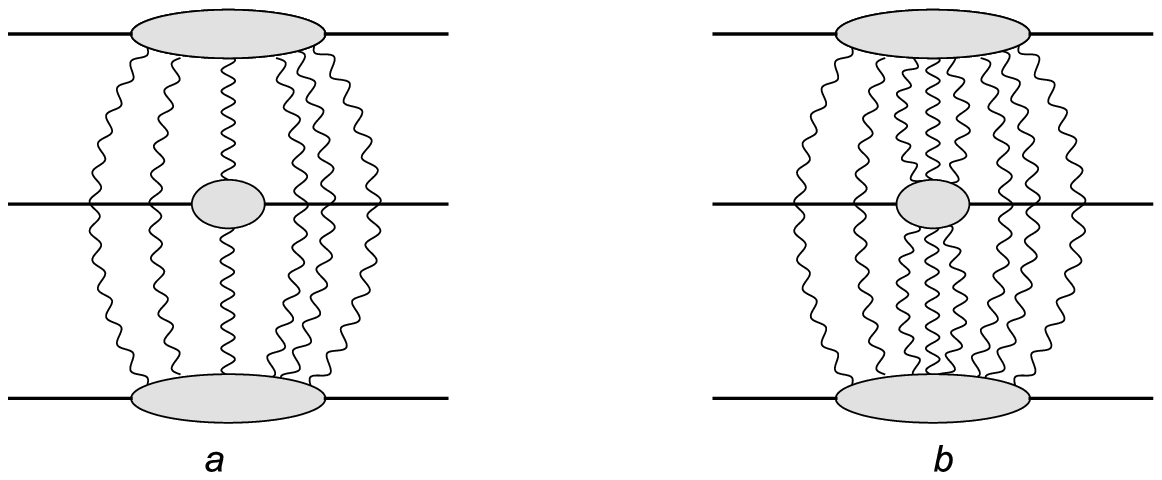}
\caption{Multipomeron exchange diagrams for one particle inclusive production, $(a)$ - diagrams calculated in \cite{Ter-Mart}, $(b)$ - diagrams calculated in \cite{Mart85,Likh90}}
\label{fig:multipom1}
\end{center}
\end{minipage}
\hspace{2.cm}
\noindent
\begin{minipage}{7.cm}
\begin{center}
\includegraphics[scale=0.6]{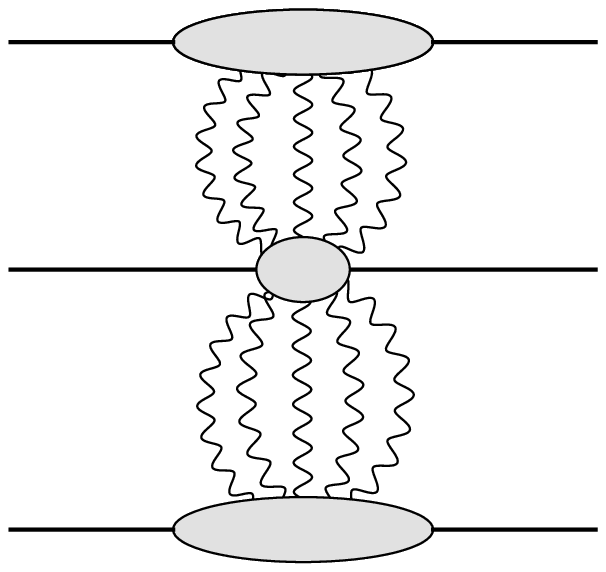}
\caption{The dominating  contributions to the central inclusive production at $s\to \infty$}
\label{fig:incl-c-pom}
\end{center}
\end{minipage}
\end{figure}

In the \cite{Ter-Mart}
the contribution to inclusive cross section of the diagrams given on the Fig. \ref{fig:multipom1},a have been calculated. It was shown that due to Abramovsky-Gribov-Kancheli rules \cite{AGK} only input diagram with one pomeron exchange contributes, rest sum of diagrams vanishes. Thus the cross section is determined by Eq.(\ref{eq:incl-c-s-y}).

Making use of the sum rule
\begin{equation}\label{eq:sumrule}
\int\frac{d^{3}p}{E}E\frac{d^{3}\sigma (ab\to cX)}{d^{3}p}=<n_{c}>\sigma_{t}^{ab}(s)
\end{equation}
and taking into account that $\sigma_{t}\approx \sigma_{0}\ln^{2}(s/s_{0})$ at $s\to \infty$ one can find
\begin{equation}\label{eq:distrib-s}
\frac{dn_{c}}{dy}=\frac{1}{\sigma_{t}(s)}8\pi g_{a}(0)\tilde V_{c}g_{b}(0)(s/s_{0})^{\varepsilon}\propto
\frac{(s/s_{0})^{\varepsilon}}{\ln^{2}(s/s_{0})},
\end{equation}
where $\tilde V_{c}$ is the integral of $v_{c}(p_{t}^{2})$ over $p_{t}^{2}$.
Then integrating Eq.(\ref{eq:distrib-s}) over $y$ we obtain a power growth for mean multiplicity, $<n_{c}>\propto (s/s_{0})^{\varepsilon}/\ln(s/s_{0})$. Let us note that $dn_{c}/dy$ does not depend on $y$ that is not supported by experimental data.

However, more diagrams must be added to calculate the inclusive cross section under interest. These diagrams are shown on Fig. \ref{fig:multipom1},b and were calculated in \cite{Mart85,Likh90}. Likewise  the case of diagrams Fig. \ref{fig:multipom1},a the sum of all contributions with the reggeons between hadrons $a$ and $b$ vanishes (see Appendix for details).

Finally, due to the above mentioned cancelation at $s\to \infty$ the inclusive cross section in the central region is dominated by contribution of the diagrams of the Fig. \ref{fig:incl-c-pom} and can be written in a general form as
\begin{equation}\label{eq:incl-c-gen}
E\frac{d^{3}\sigma}{d^{3}p}=g_{a}(0){\cal F}(y_{0}-y)v_{c}(p_{t}^{2}){\cal F}(y_{0}+y)g_{b}(0).
\end{equation}
If the input pomeron is simple $j$-pole and has intercept $\alpha_{P}(0)=1+\varepsilon$ then
\begin{equation}\label{eq:gen-2}
{\cal F}(y_{0}\pm y)=(y_{0}\pm y)^{2}.
\end{equation}
It is necessary to note that this result exactly coincide with those which can be obtained if we assume from the beginning that pomeron at $t=0$ is the triple $j$-pole at $j=1$.

This fact and similar ones valid for elastic amplitude give us some ground for a more general assumptions and allows to consider the diagram of Fig. \ref{fig:one-Pom} with pomerons of arbitrary hardness at $t=0$.

If pomeron contribution to partial amplitude (of elastic scattering) at $t=0$ is proportional to $1/(j-1)^{\nu +1}$ ($\nu \leq 2$ because of the unitarity bound) then one can argue that in Eq.(\ref{eq:incl-c-gen})
\begin{equation}\label{eq:gen-nu}
{\cal F}(y_{0}\pm y)=(y_{0}\pm y)^{\nu}\qquad  \mbox{\rm and}\qquad \frac{dn}{dy}\propto \frac{1}{\sigma_{t}(s)}(y_{0}-y)^{\nu}(y_{0}+y)^{\nu}.
\end{equation}
We would like to remark that such a behaviour of $dn/dy$ (at $\nu >0$) is in a qualitative agreement with high energy experimental data, which show a rise $dn/dy$ at $y_{0}$ and a behaviour in $y$ closed to parabolic form. Taking into account that such a pomeron leads to $\sigma_{t}(s)\propto \ln^{\nu}(s/s_{0})\approx y_{0}^{\nu}$ one can find from Eq.(\ref{eq:sumrule}) at $s\to \infty$.
\begin{equation}\label{eq:mplic-gen}
\frac{dn}{dy}(y=0)\propto \ln^{\nu}(s/s_{0})\quad {\rm and}\quad <n>\,\, \propto \,\ln^{1+\nu}(s/s_{0}).
\end{equation}

It is known (see {\it e.g.} \cite{meanmultexp}) that a good description of the mean hadron multiplicity is achieved within a logarithmic energy dependence with $\nu=1$ (dipole pomeron) or $\nu=2$ (tripole pomeron). The above mentioned properties of the unitarized pomeron models concerning of one particle inclusive distribution are rather attractive, therefore they should be checked out quantitatively with the data. We do that in the next section.

\section{Comparison of the unitarized pomeron models with the data}

\subsection{Experimental data}

\begin{figure}[h]
\begin{minipage}[t]{6.5cm}
\begin{center}
\includegraphics[scale=0.6]{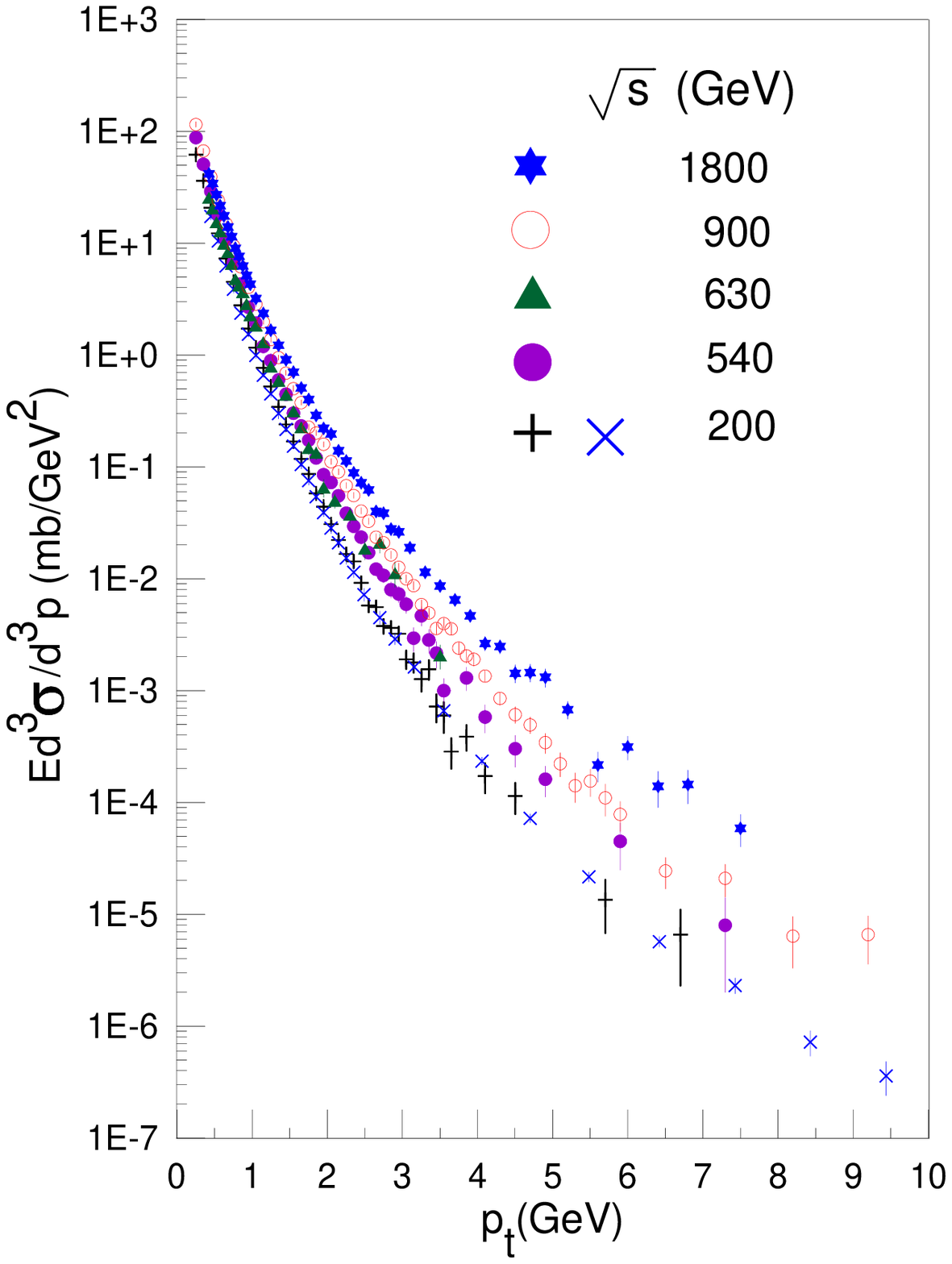}
\caption{Invariant cross section of charged hadrons
 $\bar pp\to {\rm charged}+X$, $Ed^{3}\sigma/d^{3}p$. Data are taken from \cite{difcs,UA1-1990,dnde}}
\label{fig:d3sd3p}
\end{center}
\end{minipage}
\hspace{2.cm}
\begin{minipage}[t]{7.6cm}
Even for the chosen high energies we see a nontrivial dependence of cross sections on $p_{t}$. One can clearly see that slope is changing with energy. Taking into account that $p_{t}$-dependence in the pomeron contribution is coming only from vertex function  $v_{c}(p_{n}^{2})$, one can expect that the slope effect can be explained in the model only due to subasymptotic contributions. For example, in the dipole model it can be simple pole contribution with $\alpha_{P}(0)=1$ as well as $f$-reggeon contributing at not highest energies. Another important feature of the data is a changing of an exponential increasing $Ed^{3}\sigma/d^{3}p$ at small transverse momenta $p_{t}<1$ GeV for a power like increasing at higher  $p_{t}$ (larger than 1 Gev). This transition can be parameterized in appropriate way.
The data on the Fig. \ref{fig:d3sd3p} are known for $<y>=0$ while the models should be valid at an arbitrary $y$ in the central region.

Another set of the data, namely, $dn/d\eta$ is more interesting and important for our aim. This observable can be obtained be integration of $Ed^{3}\sigma/d^{3}p$ over $p_{t}$ and with a transformation from $y$ to $\eta$.
\end{minipage}
\end{figure}

Our aim is not the detailed description of all data, we would like  to demonstrate only a possibility of the considered models to reproduce the main features of the high energy data.
Evidently, at lower energy we need to add more Regge contributions which increase the number of the fitting parameters.
To avoid an extra number of contributions and parameters  we deal with the data on $Ed^{3}\sigma/d^{3}p$ at $\sqrt{s}=$ 200, 540, 630, 900, 1800 GeV (240 points) \cite{difcs,UA1-1990}
and on $dn/d\eta$ \cite{difcs,dnde} normalized to $\sigma_{in}$ (48 points). The data are shown on Fig.(\ref{fig:d3sd3p}) 

\begin{equation}\label{eq:dndeta}
\frac{dn}{d\eta}=\frac{2\pi}{\sigma_{in}}\int\limits_{0}^{p_{t,max}} \frac{d^{2}\sigma}{d^{2}p_{t}d\eta}p_{t}dp_{t}=\frac{2\pi}{\sigma_{in}}\int\limits_{0}^{p_{t,max}}
\sqrt{1-\frac{m^{2}}{m_{t}^{2}\cosh^{2}y}}\frac{d^{2}\sigma}{d^{2}p_{t}dy}p_{t}dp_{t}
\end{equation}
where $\sigma_{in}$ is inelastic cross section (we fitted the data on $dn/d\eta$, presented in \cite{difcs,UA1-1990,dnde} just for for this case). In order to perform the integration one has to know the various vertex functions $v_{c}(p_{n}^{2})$ in $d^{2}\sigma/d^{2}p_{t}dy$ which are not determined within any Regge model. Therefore we parameterize them in a some form just to reproduce the experimental data. We would like to emphasize here that the explicit form of $p_{t}$-dependence is not crucial for models under interest. It plays only a subsidiary role in obtaining $d\sigma/d\eta$ or $dn/d\eta$. Dependence of the differential cross section on $y$ is more important for a verification of our approach.

\subsection{Double pomeron pole (or dipole)}
At $\sqrt{s}\geq 200$ GeV we take into account  dipole (d) and simple (p) poles with $\alpha(0)=1$ and $f$-reggeon for the both reggeons (upper and lower) in the diagram of Fig. \ref{fig:one-Pom}. Thus, we have six terms for cross section
\begin{equation}\label{eq:dipole-mod}
\begin{array}{ll}
{\displaystyle E\frac{d^{3}\sigma}{d^{3}p}}=& g_{dd}v_{dd}(p_{t})(y_{0}-y)(y_{0}+y)+2y_{0}g_{dp}v_{dp}(p_{t})+
g_{pp}v_{pp}(p_{t})+\\
&g_{df}v_{df}(p_{t})[(y_{0}-y)e^{\varepsilon_{f}(y_{0}+y)}+ (y_{0}+y)e^{\varepsilon_{f}(y_{0}-y)}]+\\
& g_{pf}v_{pf}(p_{t})[e^{\varepsilon_{f}(y_{0}-y)}+e^{\varepsilon_{f}(y_{0}+y)}]+
g_{ff}v_{ff}(p_{t})e^{2\varepsilon_{f}y_{0}}
\end{array}
\end{equation}
where $g_{ab}, a,b=d,p,f$ are constants, $\varepsilon_{f} =\alpha_{f}(0)-1$, $v_{a,b}(p_{t})$ are vertex functions defined as follows
\begin{equation}\label{eq:vertex}
v_{ab}(p_{t})=(1+p_{t}^{2}/p_{0}^{2})^{-\mu_{ab}}[e^{-Bp_{t}}+c(1+p_{t}^{2}/p_{1}^{2})^{-\mu}].
\end{equation}

\subsection{Triple pomeron pole (or tripole)}
Generally there are four terms to be accounted for in this model. They are triple (t), double (d) and simple (p) pomeron poles together with $f$-reggeon. In total we have 12 terms including interferences. To avoid too many parameters we consider here a more simple model which has the form
\begin{equation}\label{eq:tripole-mod}
\begin{array}{ll}
{\displaystyle E\frac{d^{3}\sigma}{d^{3}p}}=& g_{tt}v_{tt}(p_{t})z_{u}z_{d}+g_{tp}v_{tp}(p_{t})(z_{u}+z_{d})+g_{pp}v_{pp}(p_{t})+\\
&g_{tf}v_{tf}(p_{t})[z_{u}e^{\varepsilon_{f}(y_{0}+y)}+ z_{d}e^{\varepsilon_{f}(y_{0}-y)}]+\\
& g_{pf}v_{pf}(p_{t})[e^{\varepsilon_{f}(y_{0}-y)}+e^{\varepsilon_{f}(y_{0}+y)}]+
g_{ff}v_{ff}(p_{t})e^{2\varepsilon_{f}y_{0}}
\end{array}
\end{equation}
where $t$- and $d$-terms are combined to the single term as 
$$
z_{u}=\beta (y_{0}-y)^{2}+ (y_{0}-y),\qquad
z_{d}=\beta (y_{0}+y)^{2}+ (y_{0}+y)
$$
and functions $v_{ab}(p_{t})$ are defined by Eq.(\ref{eq:vertex}).

\subsection{Simple pomeron pole}
This model (violating unitarity bound) we consider just to compare with the previous ones. We take into account two pomeron poles: one (``supercritical'') has intercept $\alpha_{P}(0)=1+\varepsilon$, another has $\alpha(0)=1$.
\begin{equation}\label{eq:dipole-mod}
\begin{array}{ll}
{\displaystyle E\frac{d^{3}\sigma}{d^{3}p}}=& g_{ss}v_{ss}(p_{t})s_{u}s_{d}+g_{sp}v_{sp}(p_{t})(s_{u}+s_{d})+g_{pp}v_{pp}(p_{t})+\\
&g_{sf}v_{sf}(p_{t})[s_{u}e^{\varepsilon_{f}(y_{0}+y)}+ s_{d}e^{\varepsilon_{f}(y_{0}-y)}]+\\
& g_{pf}v_{pf}(p_{t})[e^{\varepsilon_{f}(y_{0}-y)}+e^{\varepsilon_{f}(y_{0}+y)}]+
g_{ff}v_{ff}(p_{t})e^{2\varepsilon_{f}y_{0}}
\end{array}
\end{equation}
with
$s_{u}=e^{\varepsilon(y_{0}-y)},\quad s_{d}=e^{\varepsilon(y_{0}+y)}$ and $v_{ab}(p_{t})$ taken from Eq.(\ref{eq:vertex}).

\subsection{The data fit}

\begin{figure}[ht]
\begin{minipage}[t]{7cm}
\includegraphics[scale=0.6]{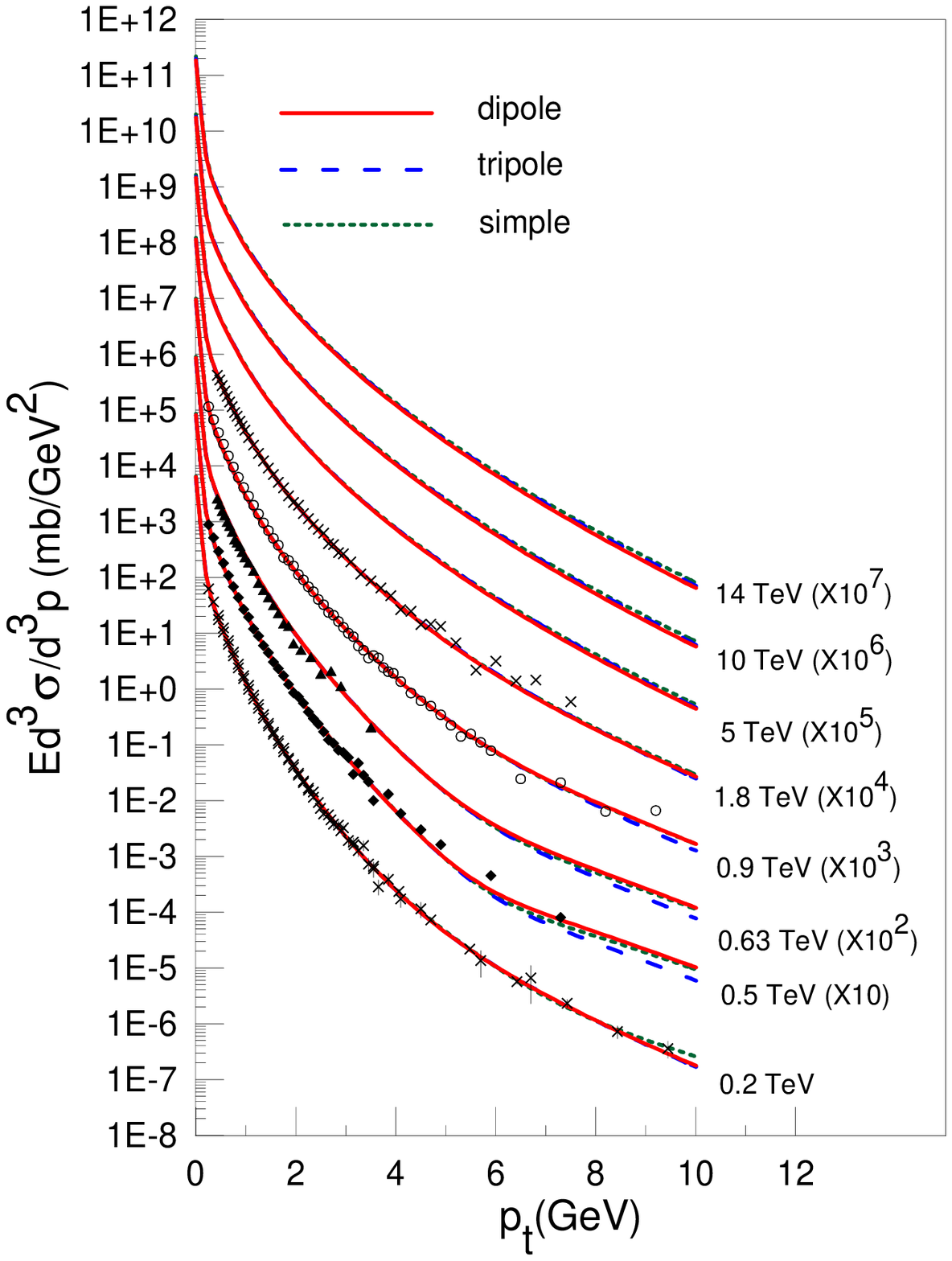}
\caption{$p_t$-dependence of inclusive cross sections at high energies. Data are taken from \cite{difcs,UA1-1990}. Curves correspond to considered models. Red solid line - dipole pomeron model, blue long dashed line - tripole pomeron model, green doted line - simple pomeron pole with $\alpha(0)>1$. Predictions for three LHC energies are shown as well.}
\label{fig:pt-distrib}
\end{minipage}
\hspace{1.cm}
\begin{minipage}[t]{7.5cm}
\includegraphics[scale=0.6]{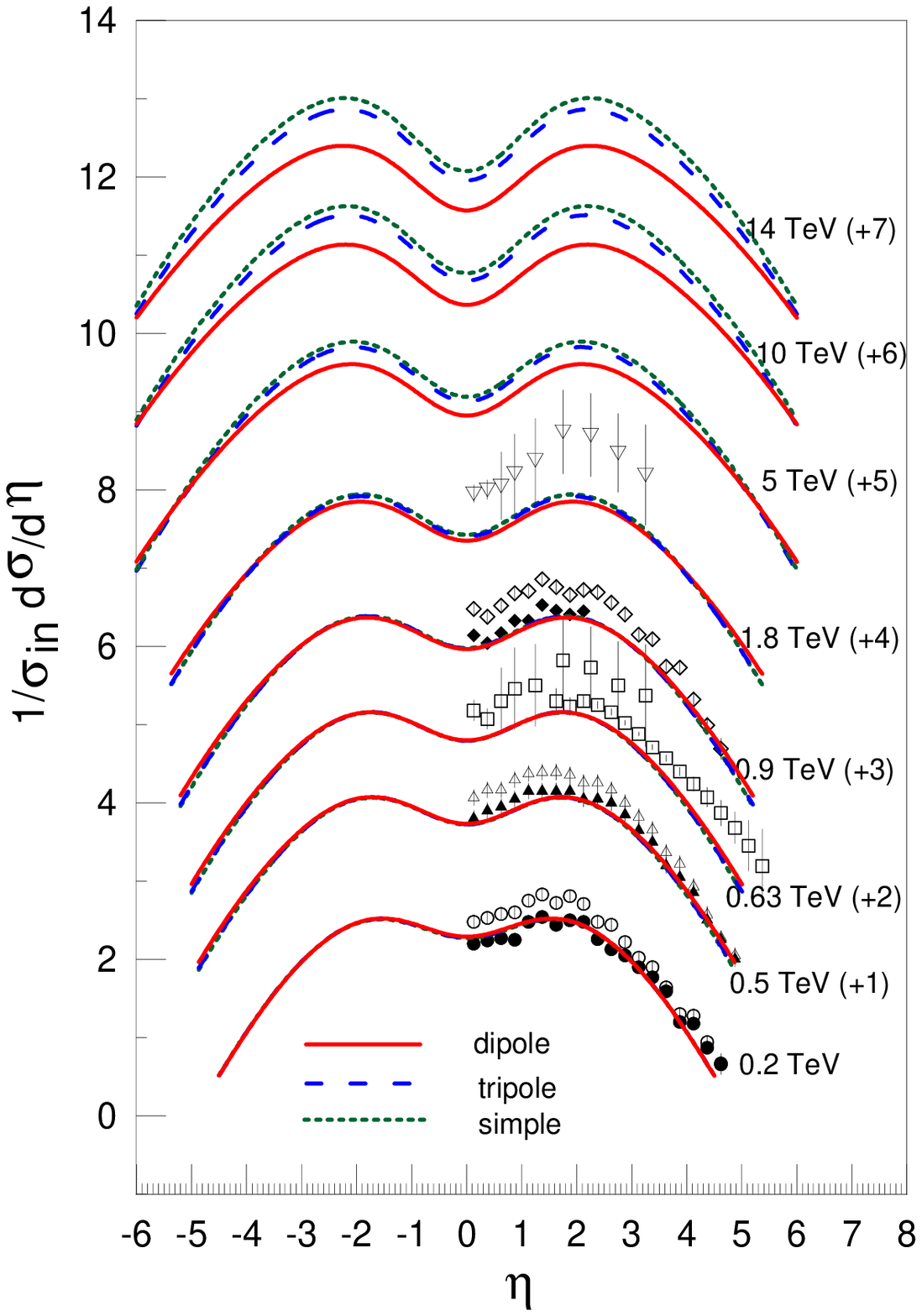}
\caption{Density of the produced hadrons as function of pseudorapidity and energy. Red (solid), blue (dashed) and green (dotted) lines are theoretical values correspondingly in dipole, tripole and simple pole pomeron model. Solid symbols correspond the data normalized to $\sigma_{in}$, open symbols correspond to data normalized to $\sigma_{NSD}$ (not used in the fit procedure). Predictions for three LHC energies are shown as well.}
\label{fig:dndeta}
\end{minipage}
\end{figure}

Parameters of the models as well as $\chi^{2}$ obtained in the fits are given in the Table, description of the data is demonstrated on Figs. \ref{fig:pt-distrib} and \ref{fig:dndeta}.

{\small
\begin{table}[h]
  \centering
  \caption{Parameters of the models obtained from the data fit}\label{tab:parameters}
\begin{tabular}{|lr|c|c|c|}
\hline
  Parameters& & Dipole & Tripole & Simple pole \\
\hline
 $g_{dd}, g_{tt}, g_{ss}$ & (mb)$^{1/2}$ & $2.93\pm 0.006$ & 1.519$\pm$0.004 & 817.9$\pm0.2$ \\
\hline
 $g_{dp}, g_{tp}, g_{sp}$ & (mb)$^{1/2}$ & $-11.38\pm 0.018$ & -8.061$\pm$0.019 & -1005.8$\pm$0.15 \\
\hline
 $g_{pp}$ & (mb)$^{1/2}$ & 61.57$\pm$0.56 & $79.54\pm 0.56 $ & 1265.9$\pm$0.46 \\
\hline
 $g_{df}, g_{tf}, g_{sf}$ & (mb)$^{1/2}$ & $0.61\pm 0.002$ & 1.002$\pm$0.002 & 5.106$\pm$0.005 \\
\hline
 $g_{pf}$ & (mb)$^{1/2}$ & -7.54$\pm$0.023) & $-28.39\pm 0.03$ & -12.283$\pm$0.01\\
\hline
 $g_{ff}$ & (mb)$^{1/2}$ & 1175.9$\pm$7.2 & $272.62\pm 0.50$ & 260.51$\pm$0.67 \\
\hline
  $\beta$ &  & - & $0.031\pm 0.001$ & - \\
\hline
  $\varepsilon $ & & - & - & 0.046 ($<0.001$) \\
\hline
  $\varepsilon_{f} $ & & 0.305$\pm$0.005 & 0.201 ($<$0.001) & 0.2$\pm$0.001 \\
\hline
  $p_{0}$ & GeV & 0.525$\pm$0.002 & 0.518$\pm$0.002 & 0.492$\pm$0.002 \\
\hline
 $p_{1}$ & GeV & 0.070$\pm$0.001 & 0.073$\pm$0.001) & 0.072$\pm$0.001\\
\hline
 B & GeV$^{-1}$ & 0.864$\pm$0.009 & 0.931$\pm$0.008 & 0.895$\pm$0.007 \\
\hline
 $A$ & & 83.52$\pm$1.91 & 82.85$\pm$1.80 & 82.87$\pm$1.63 \\
\hline
 $\mu$ & & 3.828$\pm$0.037 & 3.872$\pm$0.037 & 3.854$\pm$0.035  \\
\hline
 $\mu_{dd}, \mu_{tt}, \mu_{ss}$  & & 1.349$\pm$0.001 & 1.352$\pm$0.001 & 1.370 ($<0.001$) \\
\hline
  $\mu_{dp}, \mu_{tp}, \mu_{sp} $ & & 1.289$\pm$0.001 & 1.313$\pm$0.001 & 1.378 ($<$0.001) \\
\hline
   $\mu_{pp}$ & & 2.066$\pm$0.012 & 1.818$\pm$0.091 & 1.417$\pm$0.001 \\
\hline
  $\mu_{df}, \mu_{tf}, \mu_{sf}$ & & 0.881$\pm$0.001 & 0.388 ($<$0.001) & 0.257$\pm$0.002 \\
\hline
  $\mu_{pf}$ & & 0.329$\pm$0.001 & 0.469 ($<$0.001) & 0.314 ($<$0.001) \\
\hline
  $\mu_{ff}$ & & 0.881$\pm$0.001 & 0.597 ($<$0.001) & 0.713$\pm$0.001 \\
\hline
  $\chi^{2}_{tot}$ & & 795.9 & 809.4 &  787.1  \\
\hline
  $\chi^{2}/n.d.f.$ & & 3.01 & 2.98 & 2.96 \\
\hline
\end{tabular}
\end{table}
}

\begin{figure}[h]
\begin{minipage}[t]{6.5cm}
\begin{center}
\includegraphics[scale=0.4]{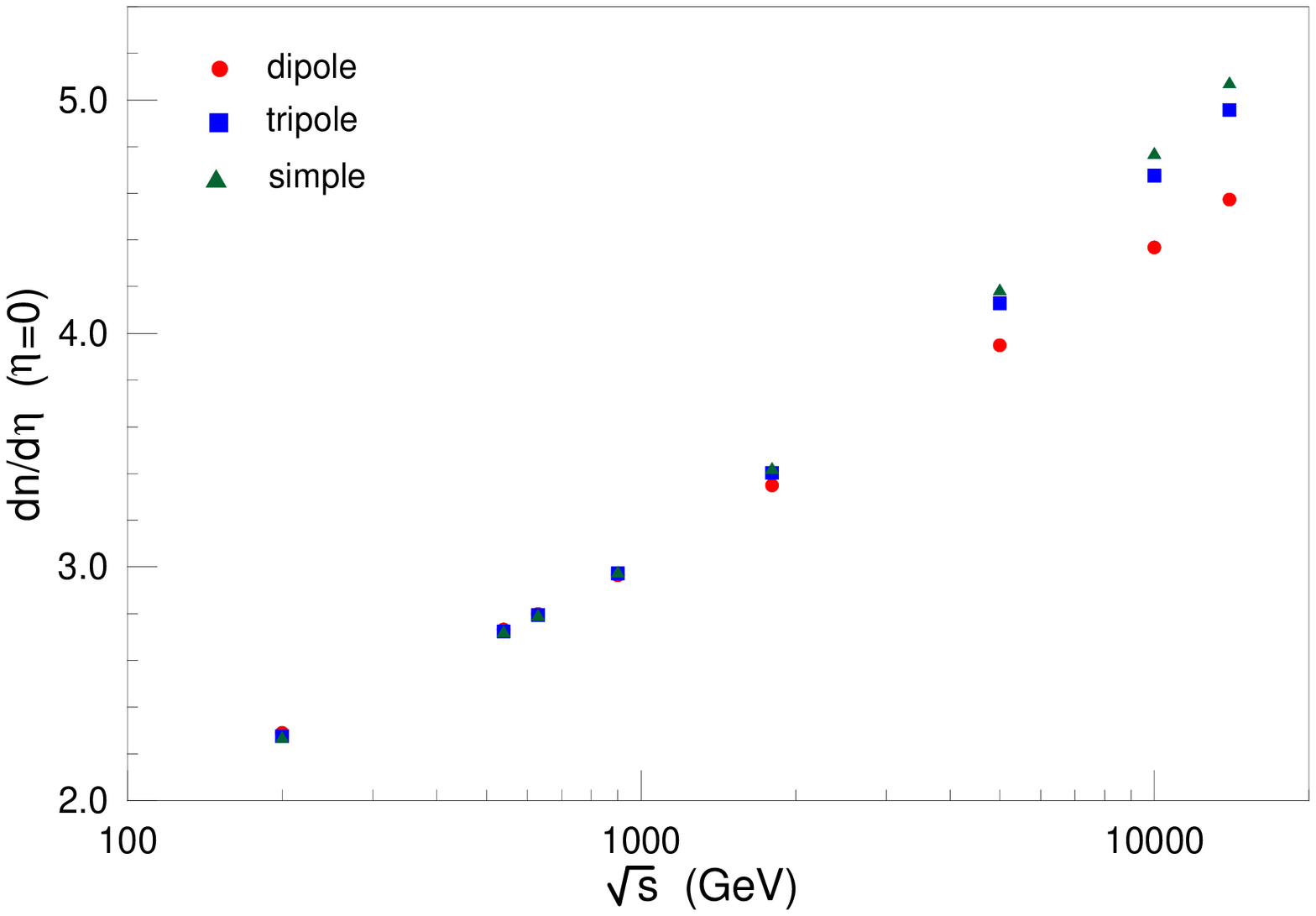}
\caption{Density of the produced hadrons at $\eta =0$ as function of energy. }
\label{fig:dndeta0}
\end{center}
\end{minipage}
\hspace{1.cm}
\begin{minipage}[t]{6.5cm}
\begin{center}
\includegraphics[scale=0.4]{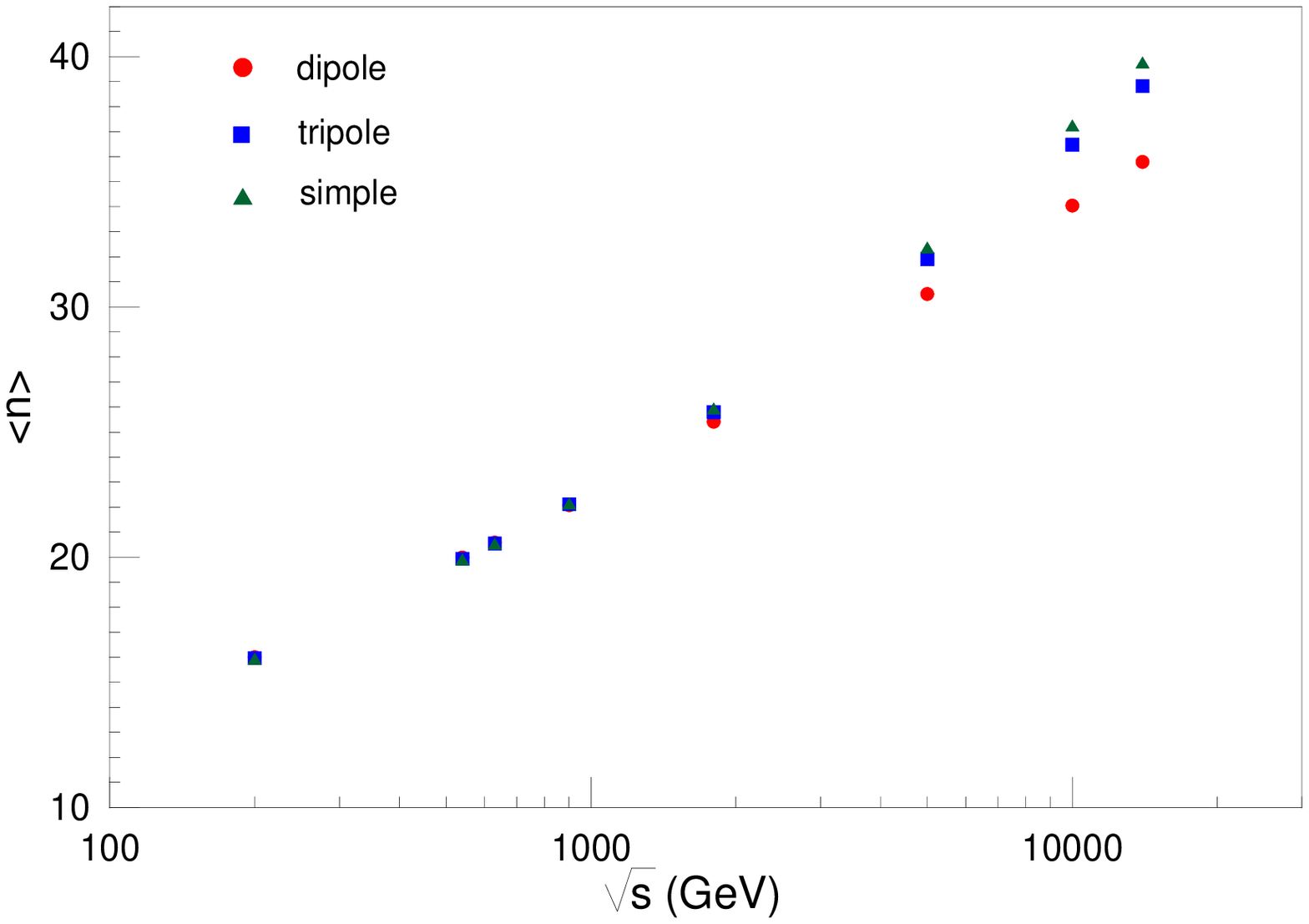}
\caption{Mean multiplicity of the produced hadrons as function of energy, calculated in the interval $-3.5\leq \eta \leq 3.5$.}
\label{fig:meanmult}
\end{center}
\end{minipage}
\end{figure}

\begin{figure}[h]
\begin{center}
\includegraphics[scale=0.4]{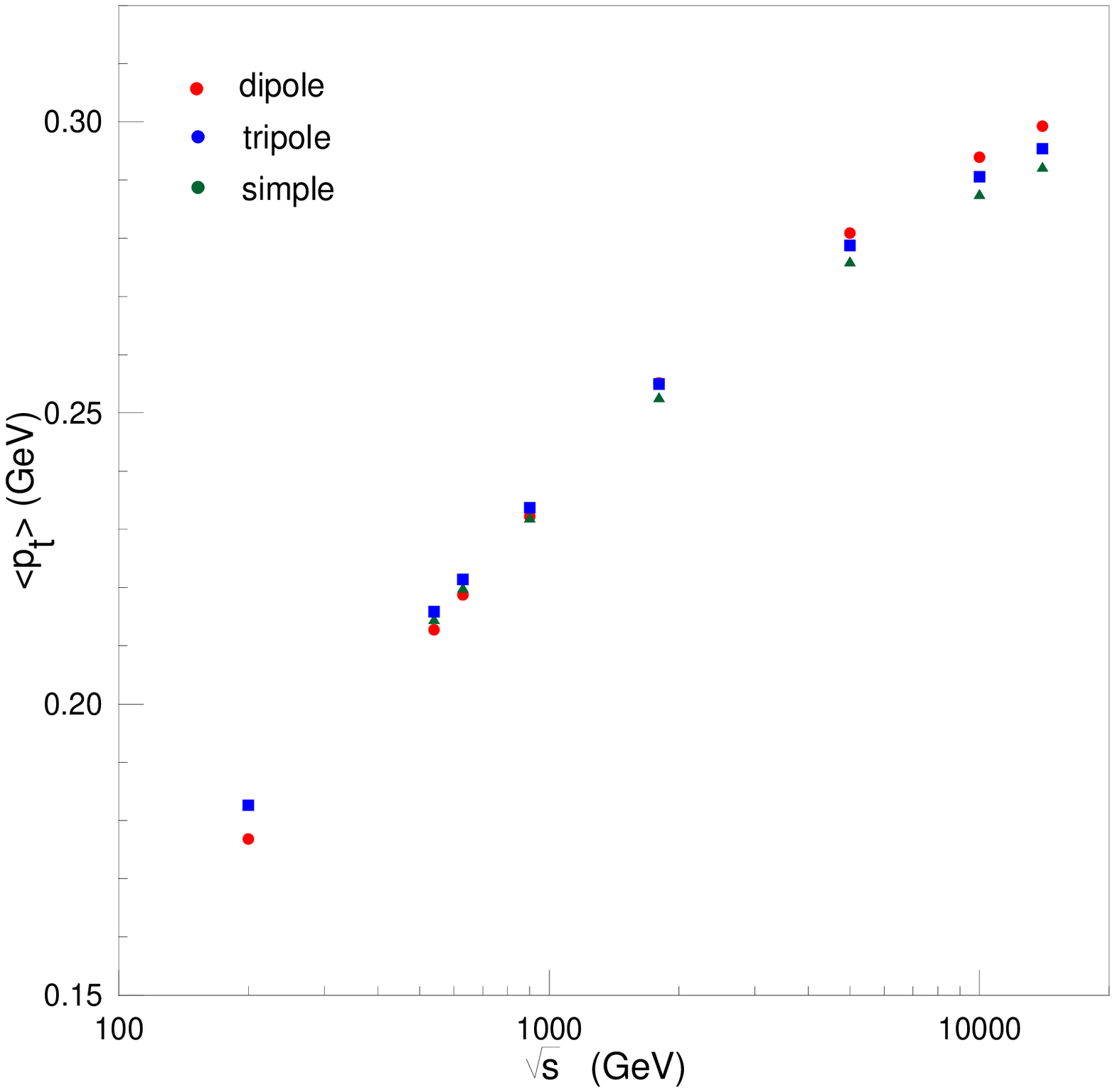}
\caption{Average transverse momentum of the produced hadrons as function of energy. }
\label{fig:aver-pt}
\end{center}
\end{figure}

One can see that theoretical curves in three models are very close each to other, at least for energies where data exist. It is not a surprise because the parameters of the tripole and simple models in fact mimic the dipole pomeron model. In the tripole model parameter $\beta$ is equal to 0.03 thus the terms containing $(y-y_{0})^{2}$ are not too important at the achieved energy, $\sqrt{s}\leq 1800$ GeV. A similar situation occurs in the simple pomeron model where a strong cancelation among the $s$- and $p$-terms mimics a logarithmic behaviour at these energies.
$$
g_{ss}(s/s_{0})^{\varepsilon}+g_{pp}\approx g_{ss}+g_{pp}+g_{ss}\varepsilon \ln(s/s_{0})
$$
However a difference between the models' predictions is increasing with energy.
It can be seen clearly on the Figs. \ref{fig:dndeta0},\ref{fig:meanmult},\ref{fig:aver-pt} which demonstrate behaviour of $dn(\eta=0)/d\eta$, $<n>$ and $<p_{t}>$ in energy.

Concerning the average transferred momenta we would like to draw attention to the following property of the considered models. We have obtained lower values for $<p_{t}>$ than usually follows from a simple extrapolation of the data on $Ed^{3}\sigma/d^{3}p$ to $p_{t}=0$ (see for example \cite{UA1-1990}). It is caused by the quite high values of $Ed^{3}\sigma/d^{3}p$ at $p_{t}\to 0$ (Fig. \ref{fig:pt-distrib}) which are necessary to obtain after integration over $p_{t}$ the correct values of $dn/d\eta$. Thus the considered models predict a fast increasing of the one-particle inclusive distribution at very small $p_{t}$.

\section{Conclusion}
We have shown that the high energy experimental data on one-particle inclusive distribution can be described well in the models of unitarized pomeron, which do not violate unitarity Froissart bound for total cross section. The dipole (tripole) pomeron model, correspondingly lead to $dn(y=0)/d\eta \propto \ln s \quad (\ln^{2}s)$ and $<n>\propto \ln^{2}s \quad (\ln^{3}s)$. These models predict small differences in $dn(y=0)/d\eta $ and $<n>$ at low LHC energies while they are increasing with energy.

{\bf Acknowledgements}

We thank M. Poghosyan for helpful discussion.

\section{Appendix}
Contribution to the discontinuity ($\Delta \mathcal M$) in $M^{2}$ of the diagram with $m_{1}$ reggeons ($l_{1}$ of them are cut) in the upper part, with $m_{2}$ reggeons ($l_{2}$ of them are cut) in the bottom part, with $m$ reggeons ($l$ of them are cut) between hadrons $a$ and $b$ has the form
\begin{equation}\label{eq:multiP}
\Delta {\mathcal M}^{(m,m_{1},m_{2})}(s,t,u)=\sum\limits_{l=0}^{m}
\sum\limits_{l_{1}=0}^{m_{1}}\sum\limits_{l_{2}=0}^{m_{2}}\Delta {\mathcal M}^{(m,m_{1},m_{2})}_{l,l_{1},l_{2}}(s,t,u)
\end{equation}
which can be rewritten as
\begin{equation}\label{eq:multiP-1}
\begin{array} {ll}
8\pi \Delta {\mathcal M}^{(m,m_{1},m_{2})}(s,t,u)=&\sum\limits_{l=0}^{m}\sum\limits_{l_{1}=0}^{m_{1}}
\sum\limits_{l_{2}=0}^{m_{2}}(-1)^{m-l}{m\choose l}(-1)^{m_{1}-l_{1}}{m_{1}\choose l_{1}}(-1)^{m_{2}-l_{2}}{m_{2}\choose l_{2}}\times \\
&\biggl [\prod\limits_{i=1}^{m}2{\rm Im}P(|t|,\vec k_{i})\prod\limits_{i_{1}=1}^{m_{1}}2{\rm Im}P(|u|,\vec k_{i_{1}})\prod\limits_{i_{2}=1}^{m_{2}}2{\rm Im}P(s,\vec k_{i_{2}})G(\{\vec k\})\biggr]
\end{array}
\end{equation}
where $P(s,\vec k)=i(s/s_{0})^{\alpha(-\vec k^{2})-1}$ is the reggeon propagator in ($s,t$)-representation, $G(\{\vec k\})$ contains all the vertex functions and the square brackets imply that the expression inside them must be integrated over the momenta of all reggeons provided the total momentum equals zero.

Because of the relation
$$
\sum\limits_{l=0}^{m}(-1)^{m-l}{m\choose l}=0 \qquad {\rm at} \qquad m\neq 0
$$
the Exp.(\ref{eq:multiP-1}) leads to the following inclusive cross section
\begin{equation}\label{eq:multiP-2}
\begin{array} {ll}
{\displaystyle E\frac{d^{3}\sigma}{d^{3}p}}=&8\pi\Delta \sum\limits_{m_{1}}
\sum\limits_{m_{2}}{\mathcal M}^{(0,m_{1},m_{2})}(s,t,u)=\sum\limits_{m_{1}}
\sum\limits_{m_{2}}\sum\limits_{l_{1}=0}^{m_{1}}
\sum\limits_{l_{2}=0}^{m_{2}}(-1)^{m_{1}-l_{1}}{m_{1}\choose l_{1}}(-1)^{m_{2}-l_{2}}{m_{2}\choose l_{2}}\times \\
&\biggl [\prod\limits_{i_{1}=1}^{m_{1}}2{\rm Im}P(|t|,\vec k_{i_{1}})\prod\limits_{i_{2}=1}^{m_{2}}2{\rm Im}P(|u|,\vec k_{i_{2}})G^{a}_{m_{1}}(\vec k_{1},\vec k_{2},\ldots ,\vec k_{m_{1}})G^{b}_{m_{2}}(\vec k'_{1},\vec k'_{2},\ldots ,\vec k'_{m_{2}})\times \\&v^{(m_{1},m_{2})}_{c}(\vec k_{1},\ldots ,\vec k_{m_{1}},\vec k'_{1},\ldots ,\vec k'_{m_{2}},p_{t}^{2})\biggr]
\end{array}
\end{equation}
where $p_{t}$ is the transverse momentum of the inclusive particle $c$.

Supposing that the vertex function $v^{(m_{1},m_{2})}_{c}$ either does not depend on $\vec k_{i}, \vec k'_{i}$ or it has factored dependence on $\{\vec k_{i}\}$ and $\{\vec k'_{i}\}$ one can obtain for $Ed^{3}\sigma/d^{3}p$ an asymptotic behaviour similar to Eqs.(\ref{eq:incl-c-gen}),(\ref{eq:gen-2}).

\end{document}